\documentclass[prl,twocolumn,showpacs,preprintnumbers,letterpaper]{revtex4}
\usepackage{graphicx}
\usepackage{amsmath}
\usepackage{epsfig}
\usepackage{amssymb}

\begin{document}

\title{Universal properties of a trapped two-component Fermi gas at unitarity}

\author{D. Blume$^{1,2}$}
\author{J. von Stecher$^{1}$}
\author{Chris H. Greene$^{1}$}
\affiliation{$^{1}$ Department of Physics and JILA, University of
Colorado, Boulder, CO 80309-0440} \affiliation{$^{2}$Department of
Physics and Astronomy, Washington State University,
  Pullman, Washington 99164-2814}

\begin{abstract}
We treat the trapped two-component Fermi system, in which unlike fermions
interact through a two-body short-range potential having no
bound state but an infinite scattering length.
By accurately solving the Schr\"odinger equation
for
up to
$N=6$
fermions, we show that no many-body bound states exist other than those bound by the trapping potential, and we demonstrate
 unique universal properties of the system: Certain
excitation frequencies are separated by $2 \hbar \omega$, the
wavefunctions agree with analytical predictions and a virial theorem
is fulfilled. Further calculations up to $N = 30$ determine the
excitation gap, an experimentally accessible universal quantity, and it agrees
with recent predictions based on a
density functional approach.

\end{abstract}

\maketitle

Ultracold Fermi gases are pure and controllable systems where few-
and many-body properties can be studied. Experiments
now routinely convert a weakly-interacting atomic Fermi gas into a
molecular Bose-Einstein condensate by tuning the scattering length
near a Fano-Feshbach resonance.
While the limiting behaviors of this BCS-BEC crossover are well
understood, the strongly-interacting regime remains fundamental and
challenging. In particular, the
unitary regime,
characterized by a
diverging two-body $s$-wave scattering length $a_s$, has been the
subject of much experimental
and
theoretical research
in the
cold-atom community
(see Ref.~\cite{gior07} for a review and an exhaustive list
of Refs.). Of course, two-component Fermi gases at
unitarity also greatly interest the nuclear physics community.
Cold atomic Fermi gases and neutron matter with large $a_s$ are two
physical realizations of the idealized system described in the
``Bertsch problem''~\cite{bertsch}. Despite its importance, few
essentially exact calculations exist for its energy eigenstates, in
part because the system is strongly-correlated and has no small
parameter for controlled perturbative treatments.

Here, we solve the many-body Schr\"odinger equation for the trapped
unitary Fermi gas with an even or odd number of fermions by two
different numerical techniques:  a basis set expansion technique that
uses correlated Gaussians (CG) and a fixed-node diffusion Monte
Carlo (FN-DMC) method. Our energies provide much needed benchmarks
for few-body systems with short-range interactions that are
important for the atomic,
nuclear, and condensed matter physics communities. A key result of
our study is a demonstration that this system exhibits unique
universal properties resembling those of the non-interacting system.

For the class of two-body potentials that support no $s$-wave bound
state, we find that many-body bound states do not exist for
systems
with
$N \le 6$
fermions. Our calculated excited-state spectrum for
$N \le 6$
explicitly confirms, to within our numerical accuracy,
the assertion by Werner and Castin~\cite{wern06} that the excitation
spectrum consists of families of breathing-mode states separated by
$2 \hbar \omega$. These excitations are non-trivial in the sense
that they are not associated with
center-of-mass
(CM)
excitations.  We
interpret this spacing using hyperspherical coordinates, and show
that in addition to
these families of breathing-mode excited states,
other excitation frequencies exist
that
equal
non-integer
multiples of the trapping frequency. Moreover, the
wavefunctions agree with analytically determined
solutions~\cite{wern06}, whose functional form was derived
on the basis of a universality assumption. Finally, we show that a
virial theorem is
fulfilled~\cite{wern06,thomasvirialtheorem}.
This, together with our finding
that negative energy states do not exist, shows explicitly that the
entire spectrum of systems with
$N \le 6$
is universal; previously
this had only been shown explicitly for the three-body system
~\cite{wern06prl,jons02}
but its likely validity had also been speculated
for larger $N$~(see, e.g., Ref.~\cite{wern06}, and Refs. therein).
A final key
point of the present study is a calculation of the excitation gap, for
$N \le 29$.
Encouraging agreement is found with results obtained within a
density functional theory (DFT) framework~\cite{bulgac07}, and it also 
sheds light on the applicability of the local density
approximation (LDA).

Our starting point is the 
Hamiltonian $H$ for
$N$ mass-$m$ fermions under spherical
external confinement with angular trapping frequency $\omega$,
$H=T + V_{tr} + V_{int}$, where
$T= -\hbar^2/(2m)\sum_{i=1}^N \nabla_i^2$, 
$V_{tr} = \sum_{i=1}^{N}m \omega^2 \vec{r}_i^2/2$
and $V_{int} = \sum_{j=1}^{N_1} \sum_{k=1}^{N_2} V_0(r_{jk})$.
Here $N_1$ and $N_2$ denote respectively the number of spin-up and
spin-down fermions ($N=N_1+N_2$), and $\vec{r}_i$ denotes the
position vector of the $i$th atom. The purely attractive
short-range
potentials $V_0$
(identical to
those
of
Ref.~\cite{vonstechtbp})
depend
on the interparticle distance
$r_{jk}$, 
and
are
parametrized by
two parameters, the depth $d$ and the range $R_0$, where $R_0 \ll
a_{ho}$ and $a_{ho} = \sqrt{\hbar / (m \omega)}$. For a given $R_0$,
$d$ is adjusted so that the first free-space two-body $s$-wave bound
state is just about to exist (i.e., so that $|a_s|=\infty$). To
study even-odd oscillations, we set $N_1=N_2$ for even $N$, and
$N_1=N_2+1$ for odd $N$. Two 
numerical methods are applied
to solve the Schr\"odinger equation for $H$: the CG and the FN-DMC
approach.

The CG approach has previously been used to determine the energy
spectrum of the four-fermion system with short-range interactions
\cite{stec07,vonstechtbp}. Here we extend these studies up to
$N=6$, and additionally present structural properties. In our
implementation, each basis function is written as a product of the
CM
ground state and a symmetrized product
of Gaussian functions each of which depends on one of the $N(N-1)/2$
interparticle distances. The resulting states have vanishing
relative orbital angular momentum $L_{rel}$. To describe
the $N$-atom states with arbitrary $L_{rel}$ in our CG approach, we
add an extra non-interacting particle. 
The Hamiltonian matrix is evaluated analytically.
A stochastic variational approach~\cite{suzuki1998sva} 
optimizes the basis set and
convergence.

Table~\ref{tab1} summarizes a few selected
\begin{table}
\caption{\label{tab1} CG energies $E_{\nu n}$ for a Gaussian potential;
$R_0=0.01a_{ho}$ for $N=3$ and $4$, and
$0.05a_{ho}$ for $N=5$
and $6$. 
}
\begin{ruledtabular}
\begin{tabular}{l|c|lll|ll}
 $N$ & $L_{rel}$ & $E_{00}/(\hbar \omega)$ & $E_{01}/(\hbar \omega)$ &
$E_{02}/(\hbar \omega)$ & $E_{10}/(\hbar \omega)$ & $E_{20}/(\hbar
\omega)$\\ \hline
3 &0& 4.682 &6.685 & 8.688  & 7.637 &9.628 \\
3 &1& 4.275 &6.276 & 8.279  & 6.868 & 8.229 \\
4 &0& 5.028   &  7.032 & 9.039 &7.464 &8.051\\
5 &0& 8.03  &10.04 &12.06 &8.83&10.38 \\
5 &1& 7.53
&
& &9.13 \\
6 &0& 8.48 & 10.50 & 12.52  & 10.44& 11.00 \\
\end{tabular}
\end{ruledtabular}
\end{table}
total energies $E_{\nu n}$ for $N=3-6$ (throughout this work,
CM
excitations are not considered, i.e., $E_{CM}=\frac{3}{2}
\hbar \omega$).
For $N=3$ and 4, calculations for different $R_0$
indicate that the
finite range effect of the reported energies is 
$\approx 0.02\hbar \omega$~\cite{stec07,vonstechtbp}.
For $N=5$ and $6$, 
finite range effects are expected to be $\approx 0.05 \hbar \omega$.
Assignment of the quantum numbers $\nu$ and $n$
is discussed below.
Our $N=3$ energies
agree with those reported in
Ref.~\cite{wern06prl}
and the corresponding $s_{\nu}$ coefficients (defined below)
agree with those reported in Ref.~\cite{dinc05}.
For $N=5$ and 6, our energies are the first {\em{ab initio}}
results obtained for this Hamiltonian by a method
that is at least in principle free of
any assumptions.
We find that the systems with $N=5$ ($L_{rel}=0$ and $1$)
and $N=6$
($L_{rel}=0$) support no negative energy states.

To treat up to $N=30$ fermions, the Schr\"odinger
equation is solved by the FN-DMC method~\cite{reyn82,hamm94}.
The proper fermionic
antisymmetry is imposed through the use of a guiding function
$\psi_T$. To within statistical uncertainties, the FN-DMC algorithm
provides an upper bound to the exact energy, i.e., to the
lowest-lying state with the same symmetry as $\psi_T$. Following
Ref.~\cite{vonstechtbp}, we consider two different functional forms
for $\psi_T$: The nodal surface of $\psi_{T1}$ is constructed by
antisymmetrizing a product of pair functions~\cite{astr04c}, and
that of $\psi_{T2}$ coincides with the nodal surface of the
non-interacting Fermi gas. Odd $N$ systems are studied using
generalizations of the $\psi_T$ defined in Ref.~\cite{vonstechtbp}
for even $N$: We find that $\psi_{T2}$ gives the lowest energy up to
$N \lesssim 9$ and $\psi_{T1}$ for larger $N$. Unlike the energy,
the unbiased calculation  by the FN-DMC approach of observables
associated with operators that do not commute with the Hamiltonian
is more involved. Here, we use the mixed estimator $\langle A
\rangle_{mixed}$, $\langle A \rangle_{mixed} = 2 \langle A
\rangle_{DMC}-\langle A \rangle_{VMC}$~\cite{hamm94}. 
To improve $\psi_T$, we introduce
additional two-body correlations~\cite{astr05,chan05}.

Table~\ref{tab2} summarizes the ground state energies $E_{00}$
obtained by the FN-DMC approach for $N \le 30$ (the energies for
even $N$, $N \le 20$, were already presented in Ref.~\cite{vonstechtbp}).
\begin{table}
\caption{\label{tab2} FN-DMC energies $E_{00}$ and expectation
values $2 \langle V_{tr} \rangle$, both in units of $\hbar \omega$.
The range $R_0$ of the square-well potential is
$0.01a_{ho}$ for $N \le 20$, and a bit smaller for larger
$N$. Statistical uncertainties of $E_{00}$ are reported in round brackets
and of
$2\langle V_{tr}\rangle$ are a few times larger
than those of $E_{00}$.}
\begin{ruledtabular}
\begin{tabular}{cll|cll|cl}
$N$ & $E_{00}$ & $2 \langle V_{tr} \rangle$ &
$N$ & $E_{00}$ & $2 \langle V_{tr} \rangle$ &
$N$ & $E_{00}$
 \\ \hline
3 & 4.281(4) &4.255& 13 & 24.79(9) &24.99 & 23 & 51.01(18)\\
4 & 5.051(9) &5.028& 14 & 25.92(5) &26.21 & 24 & 52.62(20)\\
5 & 7.61(1)  &7.48 & 15 & 29.59(10)&29.86 & 25 & 56.85(22)\\
6 & 8.64(3)  &8.86 & 16 & 30.88(6) &31.11 & 26 & 58.55(18)\\
7 & 11.36(2) &11.20& 17 & 34.64(12)&34.93 & 27 & 63.24(22)\\
8 & 12.58(3) &12.55& 18 & 35.96(7) &36.39 & 28 & 64.39(31)\\
9 & 15.69(1) &15.55& 19 & 39.83(15)&40.89 & 29 & 69.13(31)\\
10 & 16.80(4)&16.82& 20 & 41.30(8) &41.84 & 30 & 70.93(30)\\
11 & 20.11(7)&19.81& 21 & 45.47(15)&&  & \\
12 & 21.28(5)&21.39& 22 & 46.89(9) &&  & \\
\end{tabular}
\end{ruledtabular}
\end{table}
For $N \le 22$, our energies provide more stringent bounds with
smaller error bars than those reported previously~\cite{chan07}.
For $N=23-30$, these are the first {\em{ab initio}} results
available. Comparison with Table~\ref{tab1} shows that the FN-DMC
energies for $N \le 6$ agree with the CG energies to within 2\%.
This agreement validates our construction of the nodal surface of
$\psi_T$ for $N \le 6$.

Dilute gases with short-range interactions exhibit
unique universal properties at unitarity,
such as the predicted behavior that the
energy spectrum contains sequences of excited universal
breathing-mode states separated by
$2 \hbar \omega$~\cite{wern06,cast04}.
This unique feature, specific to the unitary
gas, can be intuitively understood by realizing that the only length
scale of the problem is 
given in this limit by the size of the system, i.e., the
hyperradius $R$.
Here,
$R$ is defined by removing the
CM vector $\vec{R}_{CM}$ and dividing the remaining $3N-3$
coordinates into the hyperradius $R$ and $3N-4$ hyperangles
$\Omega$: $m R^2=\sum_i m r_i^2-M R^2_{CM}$, where $M = N m$.
A dimensionality argument then
suggests that the functional form of the universal
hyperradial potentials $V_{s_{\nu}}(R)$ should be the same as that
of the non-interacting system,
which immediately implies the $2 \hbar \omega$ energy spacing for
any excitation operator that depends on only
$R$.

This argument has been formalized by Werner and Castin
for the $N$ particle
unitary gas
interacting through zero-range
pseudopotentials under 
harmonic confinement~\cite{wern06}:
It has been
predicted that the adiabatic approximation
is exact for universal states.
For these states,
the wavefunction
thus separates into a product of an in general unknown
$R$-independent channel function $\Phi_{\nu}(\Omega)$
and a
$\Omega$-independent radial function $F_{\nu n}(R)$, $\Psi^{rel}_{\nu
n}(R,\Omega) = R^{(4-3N)/2}F_{\nu n}(R) \Phi_{\nu}(\Omega)$.
For a given hyperangular quantum number $\nu$ ($\nu=0,1,\cdots$),
the radial quantum number $n$ takes the values $n=0,1,\cdots$.
The description of
the strongly-interacting many-body system thus reduces to solving a
one-dimensional Schr\"odinger equation for the hyperradial potential
$V_{s_{\nu}}(R)$, which includes part of the kinetic energy and a
contribution due to the 
two-body interactions,
\begin{eqnarray}
\left(\frac{-\hbar^2}{2m}\frac{d^2}{dR^2}+\frac{m \omega^2 R^2}{2}+
V_{s_{\nu}}(R) \right)F_{\nu n}
=
E_{\nu n}^{rel} F_{\nu n}
. \label{HR}
\end{eqnarray}
Here, $V_{s_{\nu}}(R)= \hbar^2 s_\nu(s_\nu+1)/(2m R^2)$. The
energies $E_{\nu n}^{rel}$ are related to the total energies $E_{\nu
n}$ through $E_{\nu n}=E_{\nu n}^{rel} + E_{CM}$. The coefficients
$s_{\nu}$ are constants that arise from the integration over the
hyperangular Schr\"odinger equation; their values are, in general,
unknown. Owing to the simple functional dependence of Eq.~(\ref{HR})
on $R$, the $E_{\nu n}^{rel}$ and
$F_{\nu n}$ can be written down readily~\cite{wern06}, $E_{\nu
n}^{rel}=(s_\nu+2n+3/2)\hbar\omega$ and $F_{\nu n}(R) =
R^{s_{\nu}+1}L_n^{(s_{\nu}+1/2)}(R^2/a_{ho}^2) \exp(-R^2/2
a_{ho}^2)$ (not normalized), where
$n$ denotes a non-negative integer, and
$L_n^{(s_{\nu}+1/2)}$ is the Laguerre polynomial. The expression for
the $E_{\nu n}^{rel}$ reveals immediately that the spacing between
energy levels within a given hyperradial potential curve
$V_{s_{\nu}}(R)$ is independent of $s_\nu$ and equals
$2\hbar\omega$.

Using the expression for $E_{\nu n}^{rel}$, the $E_{00}$ reported in
Tables~\ref{tab1} and \ref{tab2} readily give $s_0$ for $N \le 30$.
For $N \rightarrow \infty$, the ground state energy of the trapped
and homogenous systems can be connected via the LDA (see, e.g.,
Ref.~\cite{meno02}), leading to $s_0\approx\sqrt{\xi_{hom}}
E_{NI}/\hbar\omega$. Here, $E_{NI}$ denotes the energy of the
non-interacting trapped system and $\xi_{hom}$ a universal parameter
of the bulk system,
$\xi_{hom}=0.42$~\cite{carl03,astr04c,carl05,chan05}, 
which leads to
$s_0\approx0.65 E_{NI}/\hbar\omega$. 
For 
the trapped system, we
obtained $\xi_{tr}=0.465$~\cite{vonstechtbp} and $s_0\approx 0.68
E_{NI}/\hbar\omega$. 
The 
functional form of $V_{s_\nu}$
can also be obtained using a renormalization technique as
$N\rightarrow\infty$ \cite{seth07}, 
which leads to 
$s_0\approx 0.71
E_{NI}/\hbar\omega$ in fairly close agreement 
with 
the above 
values.

To date, the $2 \hbar \omega$ spacing of the
unitary gas
has only
been verified explicitly for $N=3$~\cite{wern06prl}. The spacing of
the energies $E_{\nu n}$ reported in Table~\ref{tab1} for $N=3-6$
for the 
first few
states with $\nu=0$ agrees
with the predicted $2 \hbar \omega$ spacing to better than $2$\%,
i.e., within our numerical uncertainty. We additionally verified for
$N \le 4$ that this spacing holds for the
lowest 
states
with $\nu=1$ and $2$.

\begin{figure}[h]
\includegraphics[scale=0.5,angle=0]{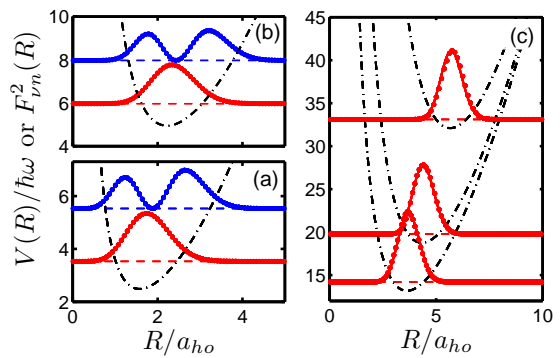}
\caption{(Color Online) Dash-dotted lines show $V(R)$ as a function
of $R/a_{ho}$ for (a) $N=4$ with $\nu=0$, (b) $N=4$ with $\nu=1$,
and (c) $N=9$, 12 and 17 (from bottom to top) with $\nu=0$. Circles
and solid lines show the corresponding $F^2_{\nu n}$ determined
numerically and analytically, respectively. Dashed horizontal lines
show the energies $E_{\nu n}^{rel}$.} \label{WF}
\end{figure}

Dash-dotted lines in Figs.~\ref{WF}(a) and (b) show the
potential curves $V(R)$, $V(R)=V_{s_{\nu}}(R)+m \omega^2 R^2/2$, for
$N=4$ for $\nu=0$ and $\nu=1$, respectively, calculated using the
$s_{\nu}$ coefficients obtained from the CG energies. Dash-dotted
lines in Fig.~\ref{WF}(c) show $V(R)$ for $\nu=0$ for $N=9$, 12 and
17 (from bottom to top) calculated using the $s_{0}$ coefficients
obtained from the FN-DMC energies. Circles and solid lines show the
corresponding square of the radial functions $F^2_{\nu n}$ obtained
numerically (by integrating 
$\Psi_{\nu n}^2$
over all coordinates but the hyperradius $R$) and analytically
(using the $s_{\nu}$ coefficients calculated from the CG and FN-DMC
energies), respectively. Clearly, our numerically determined
$F^2_{\nu n}$ agree well with those determined analytically,
providing the first quantitative numerical verification of the
interpretation of the $2 \hbar \omega$ spacing within the
hyperspherical framework for $N>3$.

Another property of the unitary Fermi gas is that all of its
\begin{figure}[h]
\includegraphics[scale=0.4,angle=0]{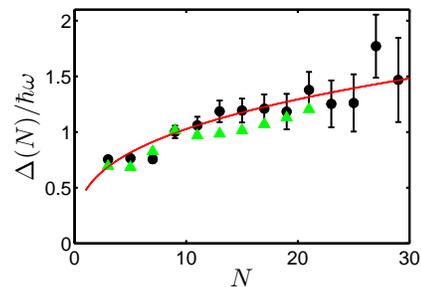}
\caption{(Color Online) Circles show 
the excitation gap 
$\Delta(N)$
determined from our FN-DMC energies. A solid line shows
$\Delta_{LDA}(N)$ using $\xi_{tr}$ and $\alpha_{tr}$. 
For
comparison, triangles show $\Delta(N)$ determined from the DFT
energies~\protect\cite{bulgac07}. } \label{fig_gap}
\end{figure}
universal states follow a virial theorem, i.e., $E_{\nu n} = 2
\langle V_{tr} \rangle_{\Psi_{\nu n}^2}$~\cite{thomasvirialtheorem,wern06}.
Table~\ref{tab2} shows that the energy expectation value (columns 2
and 5) and the energy obtained using the virial theorem (columns 3
and 6) for $N \le 20$ agree 
well, providing further
evidence for the separability of universal states of the
unitary gas
and, on the other hand, suggesting that our parametrization of
the nodal surface used in the FN-DMC calculations is adequate. The
virial theorem has previously been tested
at various levels of approximation~\cite{bulgac07,chan07,jauregui}.

We now use the ground state energies reported in Table~\ref{tab2} to
determine the experimentally measurable excitation gap $\Delta(N)$
for the trapped
unitary Fermi gas
as a function of $N$. The
definition of the excitation gap for the homogeneous
system~\cite{carl03}, which
equals half the energy it takes to break a pair, can be readily
extended to the trapped system~\cite{chan07,bulgac07},
%\begin{eqnarray}
$\Delta(N) = E_{00}(N) - [E_{00}(N-1)+E_{00}(N+1)]/2$.
%\end{eqnarray}
Circles in Fig.~\ref{fig_gap} show the excitation gap $\Delta(N)$
obtained from our FN-DMC energies. $\Delta(N)$ increases from
$\approx 0.75 \hbar \omega$ for $N=3$ to $\approx 1.5\hbar \omega$ for $N=29$.
For comparison, triangles
show
$\Delta(N)$ calculated using the DFT energies obtained recently
by Bulgac~\cite{bulgac07}. The agreement between our and Bulgac's
$\Delta(N)$ is quite good.

To gain further insight, we calculate the gap in the LDA,
$\Delta_{LDA}(N) = 3\alpha_{hom} \,(3N)^{1/3} \, \hbar
\omega/(8\sqrt{\xi_{hom}})$.
The universal parameter $\alpha$ describes the even-odd
oscillations: $\alpha_{hom}$
equals
$0.85$ for the
bulk
system~\cite{carl05}, and we find $\alpha_{tr}=0.60$
for the trapped system. $\Delta_{LDA}(N)$, shown in
Fig.~\ref{fig_gap}  by a solid line using $\xi_{tr}$ and
$\alpha_{tr}$, provides a good description of our FN-DMC results.
The fact that $\alpha_{tr}$ is noticeably smaller than
$\alpha_{hom}$ suggests that the extra particle is not distributed
uniformly throughout the cloud. Indeed, our density profiles for $N
\gtrsim 11$ (not shown) indicate that the extra particle sits near
the surface of the cloud.
In the LDA, the center should be described well but not the surface, so LDA
might fail here.
Recently, Son
proposed that the gap
increases with $N^{1/9}$
as $N \rightarrow \infty$~\cite{son07}.
A fit of our results for $\Delta(N)$
for
$N \ge 9$  shows consistency with
the $N^{1/9}$ dependence but does not conclusively confirm it.

In summary, we have shown that trapped two-component unitary Fermi gases 
with
$N \le 6$
support no negative energy
states.
While our FN-DMC calculations do not exclude the presence of
negative energy states for larger $N$, we find no evidence for their
existence. 
A 
variational
analysis
predicts an upper bound for the system size $N_{cr}$
 at which non-universal states exist of $\approx 40$.
To obtain $N_{cr}$ we consider the Gaussian two-body potential and 
take the variational many-body wave function of the untrapped system 
to coincide
with that
of a trapped non-interacting two-component Fermi gas. The confinement 
width is treated as a variational parameter, 
and $N$ is increased till a negative energy state is found.
The absence of negative energy states for
$N \le 6$ fermions explains the stability of equal-mass
two-component Fermi gases. Furthermore, we explicitly verify a
number of unique properties of the universal states of the unitary
Fermi gas.
The particular form of the energy spectrum has 
consequences:
It should be possible to verify
experimentally
that for each excitation frequency $\omega_{ex}$ there exists a
ladder of non
CM
excitations $\omega_{ex}+2 n\omega$
($n$ integer).
Furthermore, the $s_{\nu}$ coefficients calculated here for the trapped system
remain valid for the free-space system~\cite{wern06}.
Finally, we calculate the excitation gap $\Delta(N)$ for
$N \le 29$,
and find good agreement with recent DFT results. 
$\Delta(N)$ increases with $N$
and 
the excess atom
leads for $N \gtrsim 11$ to an increase of the density in the
surface region.

We acknowledge fruitful discussions with J. D'Incao, B. Esry and S.
Rittenhouse, and support by the NSF.

\end{document}